\documentclass[%
 reprint,
 amsmath,amssymb,
 aps,
 pre
]{revtex4-2}

\usepackage[hidelinks]{hyperref}% add hypertext capabilities

\usepackage{graphicx,tikz,subfigure}% Include figure files
\usepackage{dcolumn}% Align table columns on decimal point
\usepackage{bm}% bold math
\begin{document}

\preprint{APS/123-QED}

\title{Thermodynamic Bounds on the Asymmetry of Cross-Correlations  with Dynamical Activity and Entropy Production}% Force line breaks with \\

\author{Jie Gu}
\email{jiegu1989@gmail.com}
 \affiliation{Chengdu Academy of Education Sciences, Chengdu 610036, China}%Lines break automatically or can be forced with \\

\date{\today}% It is always \today, today,

\begin{abstract}
Entropy production and dynamical activity are two complementary aspects in  nonequilibrium physics. The asymmetry of cross-correlation, serving as a distinctive feature of nonequilibrium, also finds widespread utility. In this Letter, we establish two thermodynamic bounds on the normalized asymmetry of cross-correlation in terms of dynamical activity and entropy production rate. 
These bounds demonstrate broad applicability, and offer experimental testability.
\end{abstract}

\maketitle

%\tableofcontents

\textit{Introduction.---}Entropy production plays a pivotal role in nonequilibrium thermodynamics and statistical mechanics, serving as a fundamental quantity of interest \cite{landi2021}. The quest to establish connections between entropy production and measurable parameters has been a central focus in this field.
Any manifestation of nonequilibrium should be connected with entropy production or dissipation.
For instantce, maintaining instantaneous equilibrium necessitates infinitely slow driving, implying that entropy production accompanies any finite-time process. 
This  has led to the notion of thermodynamic geometry \cite{ruppeiner1995,crooks2007,sivak2012,gu2023} and speed limits in terms of entropy production \cite{shiraishi2018,vo2020,dechant2022a, nicholson2020, garcia-pintos2022, gu2023a}.
In equilibrium, currents vanish, so nonvanishing currents are also a signature of nonequilibrium, leading to the existence of bounds on entropy production in terms of nonvanishing currents normalized by its variance, known as thermodynamic uncertainty relations \cite{barato2015,gingrich2016,horowitz2020}. 
Additionally,  equilibrium conditions give rise to the principle of microscopic reversibility, implying the symmetry of cross-correlations. 
Consequently, corresponding asymmetry also serves as a distinguishing feature of nonequilibrium steady states \cite{tomita1974a}, and is  presumably associated with entropy production \cite{tomita2008}.

Recent interest has resurfaced regarding this asymmetry of cross-correlations.
Ohga \textit{et al.} have reported a fundamental inequality that explores the relationship between the normalized asymmetry of cross-correlation   and the thermodynamic forces driving the system out of equilibrium \cite{ohga2023c}.
%\begin{equation}
%\label{eq:ohga}
%	|\chi_{ab}| \equiv \frac{|\alpha_{ab}|}{\sqrt{D_a D_b}}  \leq \max _{c} \frac{\tanh \left(\mathcal{F}_{c} / 2 n_{c}\right)}{\tan \left(\pi / n_{c}\right)} \leq \max _{c} \frac{\mathcal{F}_{c}}{2 \pi},
%\end{equation}
%where  the slope of  cross-correlation asymmetry is given by $\alpha_{ab} \equiv (\partial_\tau C_{ab} -\partial_\tau C_{ba})/2$, 
%with $\partial_\tau$ denoting the derivative with respect to $\tau$ evaluated at $\tau=0$. 
%Besides,  we define the normalization term $D_a \equiv -\partial_\tau C_{aa}$, $\max_c$ the maximum over all cycles and $n_c$  the number of states in cycle $c$.
To illustrate their theory, the authors proved that the number of coherent biochemical oscillations is equivalent to the normalized asymmetry of cross-correlation between certain observables,  confirming the conjecture that the coherence of biochemical oscillations is bounded by the driving force \cite{barato2017}. 
Building upon another conjecture stating that the average entropy production per oscillation is bounded from below by the number of coherent oscillations if at least one oscillation is visible \cite{oberreiter2022}, their result also reveals a connection between asymmetry of cross-correlation and entropy production. 
Following their idea, Shiraishi found that the normalized asymmetry is bounded from above by the entropy production per characteristic maximum oscillation time \cite{shiraishi2023}.
Extensions to finite-time domain for classical \cite{liang2023} and quantum systems have been made \cite{vanvu2023b}.

On the other hand, the dynamical activity \cite{maes2006, baiesi2009a,maes2017,maes2018,maes2020} is a crucial but less explored component in nonequilibrium physics.
Its significance only emerges beyond linear order around equilibrium, and has recently been highlighted in studies on out-of-equilibrium fluctuation-response relations \cite{baiesi2009a, shiraishi2022}, classical speed limits \cite{shiraishi2018,vo2020, dechant2022a},  thermodynamic (kinetic) uncertainty relations \cite{garrahan2017, diterlizzi2019,  vo2022a,  shiraishi2023a},   inference of entropy production with lacking data \cite{harunari2022,baiesi2023},   power-efficiency trade-off in heat engines \cite{vanvu2023geometric},    upper bound on entropy production \cite{nishiyama2023},  and the thermodynamic correlation inequality \cite{hasegawa2023}.
 In essence, the dynamical activity quantifies the frequency of transitions, exhibiting a time-symmetric characteristic. In contrast, the entropy production is time-antisymmetric, changing its sign upon time reversal, thereby inverting the fluxes. 
Hence, these two quantities naturally emerge as complementary facets.
Despite its importance, the connection between the dynamical activity and entropy production remains elusive.
Complementary to previous studies \cite{ohga2023c,shiraishi2023,liang2023,vanvu2023b}, this Letter establishes two thermodynamic bounds on the normalized asymmetry of cross-correlations in terms of dynamical activity and entropy production rate [Eqs. \eqref{eq:main} and \eqref{eq:main2}].
We prove the first bound for unicyclic systems and the second bound for general cases, and present the condition for saturation.
We also provide numerical evidence to support the validity of the first bound for arbitrary network topology.
All the quantities involved are measurable, making the inequalities experimentally testable.

\textit{Setup.---} Consider a stochastic   Markov jump process with finite $\mathcal{N}$ states.
The dynamics of the probability distribution
\({\boldsymbol{p}}=[p_1,p_2,\ldots, p_N]^\mathsf{T}\) is described by the master equation \cite{kampen1992}
\begin{equation}
\label{eq:me}
    \frac{\text{d} {p}_{m}}{\text{d} t}=\sum_{n} W_{mn} p_{n},
\end{equation}
where \(p_m\) is the probability of state \(m\), \(W_{mn} (m\ne n)\) is the
time-{independent} transition rate from state \(n\) to \(m\), and the rate of leaving state $m$ is
\(W_{mm} = -\sum_{n(\ne m)} W_{nm}\).
Thermodynamic consistency is assumed, i.e., whenever $W_{mn} \ne 0$, $W_{nm}$ is nonzero too. Physically, this assumption means that the transition between two states is either forbidden or bidirectional.
We also assume that the system is in a nonequilibrium steady
state $\boldsymbol{p}^\text{ss}$ satisfying $\boldsymbol{W} \boldsymbol{p}^\text{ss} =0$.
After defining the one-way flux $\mathcal{T}_{mn } = W_{mn} p^\text{ss}_{n}$, the probability current  $\mathcal{J}_{mn}$ and local activity (also called traffic) $\mathcal{A}_{mn}$ can be defined in terms of $\mathcal{T}_{mn} (m \ne n)$ respectively as 
\begin{equation}
\label{eq:quantities}
	\begin{aligned}
&	\mathcal{J}_{mn} \equiv \mathcal{T}_{mn } - \mathcal{T}_{nm }, \\
&	\mathcal{A}_{mn} \equiv \mathcal{T}_{mn } + \mathcal{T}_{nm }.
	\end{aligned}
\end{equation}
For later use, we also define a quantity $\gamma$ that characterizes the scale separation of local activities, given by
\begin{equation}
\label{eq:gamma}
\gamma \equiv \frac{\max \mathcal{T}_{mn }}{\min \mathcal{T}_{mn }}, \quad \text{for}\,\,\mathcal{T}_{mn }>0.
\end{equation}
According to stochastic thermodynamics \cite{schnakenberg1976,seifert2012,peliti2021,shiraishi2023a}, the dynamical activity $\mathcal{A}$ and the entropy production rate $\sigma$ can be expressed as
\begin{equation}
\label{eq:daEP}
	\begin{aligned}
&	\mathcal{A} = \sum_{m \ne n} \mathcal{T}_{mn} = \sum_{m>n} \mathcal{A}_{mn},	 \\
&					\sigma  = \sum_{m\ne n} \mathcal{T}_{mn }  \ln \frac{\mathcal{T}_{mn }}{\mathcal{T}_{n m }} .
	\end{aligned}
\end{equation}

The two-time correlation between observables $a$ and $b$ at time lag $\tau$ is
\begin{equation}
C_{b a}^{\tau} \equiv \langle b(t+\tau) a(t)\rangle,
\end{equation}
where $\langle \cdot \rangle$ denotes average over trials,
and the central quantity is the normalized asymmetry of cross-correlation $\chi_{ab}$ defined as
\begin{equation}
\chi_{ab} \equiv \frac{\alpha_{ab}}{\sqrt{D_a D_b}} \equiv  \frac{(\partial_\tau C_{ba}^\tau-\partial_\tau C_{ab}^\tau)/2}{\sqrt{\partial_\tau C_{aa}^\tau \partial_\tau C_{ba}^\tau}}.
\end{equation}
The numerator $\alpha_{ab}$, i.e., the slope of cross-correlation asymmetry at $\tau=0$ (also called  stationary fluctuation oscillation in Ref. \cite{shiraishi2023}),  vanishes in equilibrium, so a nonzero asymmetry implies nonequilibrium.
The slopes of auto-correlations, $D_{aa}$ and $D_{bb}$, are a measure of diffusion \cite{ohga2023c}.
Explicitly, the slopes of cross-correlation asymmetry  and average auto-correlation can be expressed as
\begin{equation}
\label{eq:slopes}
\begin{aligned}
& \alpha_{ab}  = \frac{1}{2} \sum_{m,n}(a_n b_m -a_m b_n) \mathcal{T}_{mn} =   \sum_{m<n} \Omega_{mn} \mathcal{J}_{mn}, \\
&	\frac{{D_{a}+D_{b}}}{2} = \frac{1}{4}\sum_{m,n} [(a_m-a_n)^2+(b_m-b_n)^2] \mathcal{T}_{mn} \\
& \qquad \quad = \frac{1}{4}\sum_{m<n}  L_{mn}^{2} \mathcal{A}_{mn},
\end{aligned}
\end{equation}
with $\Omega_{mn} = (a_{n} b_{m}-a_{m} b_{n})/2$ and $L_{mn} = \sqrt{(a_m-a_n)^2+(b_m-b_n)^2}$.
Both $\Omega_{mn}$ and $L_{mn}$ have geometric meaning: $\Omega_{mn}$ is the oriented area of the triangle with $m,n$ and the origin, and $L_{mn}$ the edge between $m$ and $n$.
With all these relevant quantities, we present our main result below.

\begin{figure}[t]
    \centering
\includegraphics[width=\columnwidth]{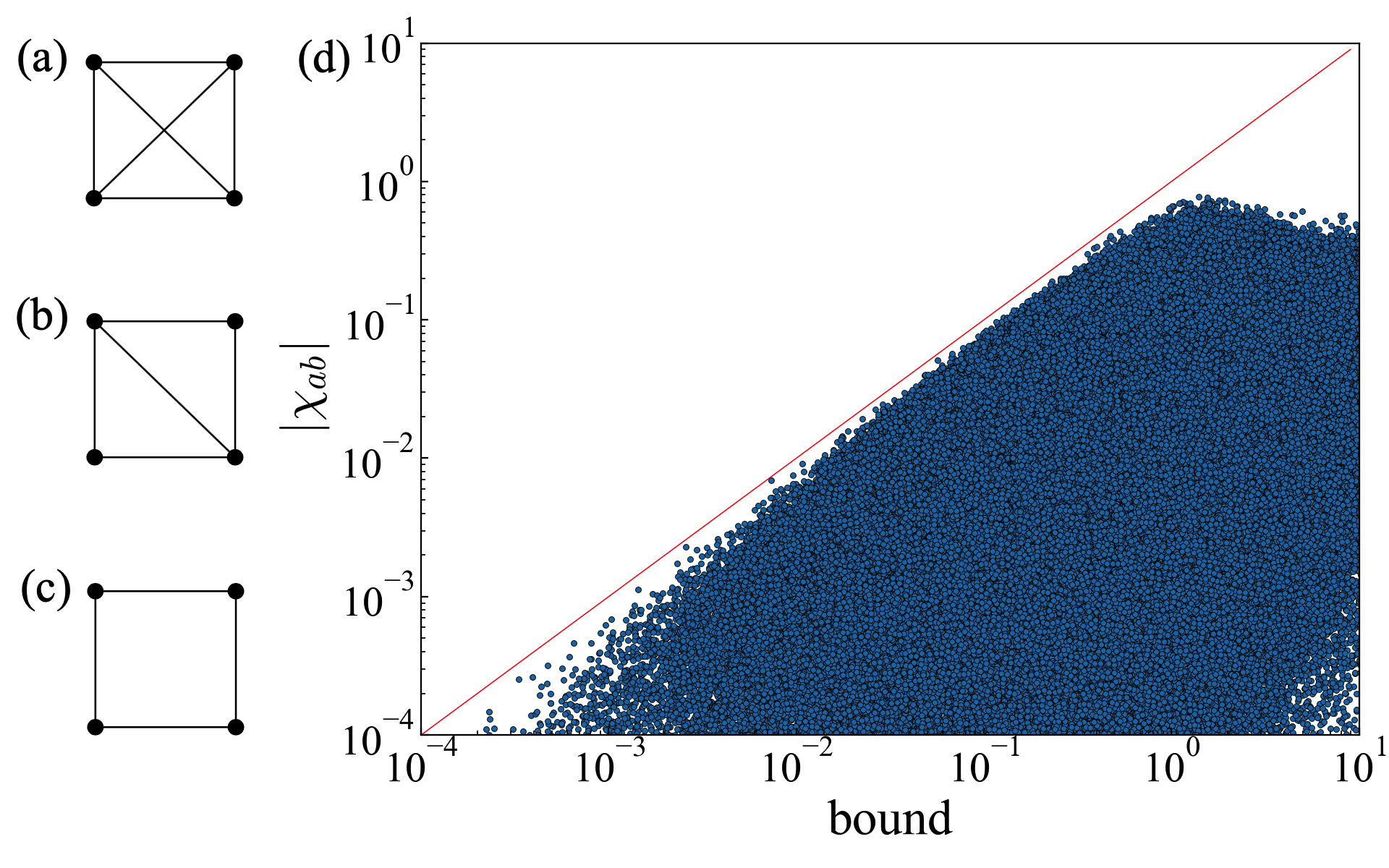}
\caption{
(a)-(c): Three topologies of networks with four states.
(d): Scatter plot of $|\chi_{ab}|$ vs. the first bound.
All data points  lie below the diagonal, which validates the bound.
\label{fig:scatter}}
\end{figure}

\textit{Main result.---}
Our main results are two thermodynamic bounds that connect the normalized asymmetry of cross-correlations $\chi_{ab}$, dynamical activity $\mathcal{A}$ and entropy production rate $\sigma$ for a Markov jump process with $\mathcal{N}$ states, given by
\begin{equation}
\label{eq:main}
	|\chi_{ab}| \le \frac{\gamma}{\tan (\pi/\mathcal{N})} \sqrt{\frac{ \sigma }{2\mathcal{A}}}
\end{equation}
and
\begin{equation}
\label{eq:main2}
	|\chi_{ab}| \le \sqrt{\frac{\mathcal{N}^*\gamma }{2\pi\tan (\pi/\mathcal{N})}} \sqrt{\frac{ \sigma }{\mathcal{A}}},
\end{equation}
where  $\mathcal{N}^*$ is the number of nonzero local activities.
The derivation of Eq. \eqref{eq:main} for unicyclic networks and Eq. \eqref{eq:main2} for general cases  is deferred to the end of the Letter.
We conjecture that Eq. \eqref{eq:main} is valid for arbitrary networks, and
present numerical evidence in Fig. \ref{fig:scatter}(d), with a total of $10^6$ data points and three distinct topologies as in Figs. \ref{fig:scatter}(a)-(c).
Each data point is generated as follows: We choose a topology randomly from the three topologies,
draw nonzero transition rates $W_{mn}$'s from the uniform distribution, calculate the diagonals of the transition rate matrix,  and randomly sample each of the observables $a_m$  and $b_m (m=1,\ldots, \mathcal{N})$ from the interval $[-1,1]$.
Subsequently,  we calculate the bound and $|\chi_{ab}|$ using Eqs. \eqref{eq:gamma}--\eqref{eq:slopes}.
The three topologies are explicitly considered because it is inherently impossible to obtain non-fully-connected networks [Figs. \ref{fig:scatter}(b)-(c)] through random sampling, but they are qualitatively different from each other.
This is because the bound is discontinuous  at $W_{mn}=0$, arising from the presence of $\gamma$.
For example, assume that $\mathcal{T}_{mn}$'s are finite for any $mn \ne 13$ and $31$, and all $p^\text{ss}_m$'s are finite as well.
If $W_{13} = W_{31} = 0$, then $\gamma = \max \mathcal{T}_{mn} / \min_{mn \ne 13, 31} \mathcal{T}_{mn}$.
However, if $W_{13}\ll 1$ and $W_{31}$ is finite, then $\gamma = \max \mathcal{T}_{mn} / \mathcal{T}_{13} $, which can be arbitrarily large.
The scatter plot clearly validates the first bound for four-state networks, and networks with a different number of states show qualitatively similar results.

It can be seen from the derivation that bound \eqref{eq:main}  saturates in unicyclic networks when the transition rates are uniform, the observables form a regular polygon, and the system is close to equilibrium. 
This saturation condition is similar to that of the thermodynamic uncertainty relation \cite{barato2015}.
In contrast, bound \eqref{eq:main2} is  saturated in unicyclic networks when the transition rates are uniform, the observables form a regular polygon, and the number of states $\mathcal{N}$ approaches infinity.
For unicyclic networks with uniform transition rates, we have $\mathcal{N}^* = \mathcal{N}$ and $\gamma =1$, and it can be proved that the first bound is tighter.
As $\mathcal{N} \to \infty$, the two bounds are asymptotically equivalent. Otherwise, either bound can be tighter depending on the parameters.

Several remarks are in order regarding the implications of our findings. 
Inequality \eqref{eq:main} reveals a thermodynamic bound on the normalized cross-correlation asymmetry, hinging upon the dynamical activity and entropy production rate---specifically, the square root of their quotient. 
This structure is natural, as will be discussed later.
The cross-correlation asymmetry  has emerged as a versatile tool widely employed to investigate an array of phenomena spanning directed interactions,  non-equilibrium oscillations, nonreciprocal motion and so on, as summarized in Ref. \cite{ohga2023c}.
Our result is applicable as long as the dynamics can be modeled by a Markov jump process, irrespective of the underlying network topology.
This includes chemical reactions, biochemical systems, and quantum transport, among others.
All the quantities constituting the bound are experimentally measurable, rendering our findings amenable to empirical validation.
Correlations can be quantified through techniques such as fluorescence cross-correlation spectroscopy  \cite{kettling1998,bacia2006}.
As for the right-hand side of the inequalities, the quantities $\gamma$ and $\mathcal{A}$ can be obtained by counting jumps in a sufficiently long trajectory, and $\sigma$ can be measured through the energetics of the environment.
Therefore, this far-from-equilibrium relation is  in principle experimentally testable.

% We admit that it is not a good inequality for inference, because $\gamma$ contains as much information as $\sigma$.

\textit{Connection and comparison with previous works.---}
We begin by comparing our findings with the seminal work of Ohga \textit{et al.} \cite{ohga2023c}. Both studies impose an upper bound on the normalized asymmetry of cross-correlation, while a crucial distinction emerges: the bound established in Ref. \cite{ohga2023c} relies on the employment of the maximum cycle affinity as a thermodynamic quantity, while our bound focuses on the entropy production rate.
The bounds presented in Refs. \cite{shiraishi2023} and \cite{vanvu2023b}   incorporate the entropy production rate as a key factor, too. 
However, a notable distinction between our result and theirs lies in the observable dependency of their bounds, while our derived bound is independent of any specific observables.
This distinction, akin to the approach employed in Ref. \cite{ohga2023c}, arises from the consideration of the normalized asymmetry,  in which the normalization factor in the denominator already encapsulates the information of observables. 
In contrast, the bounds proposed in Refs. \cite{shiraishi2023} and \cite{vanvu2023b}  pertain directly to the asymmetry itself, rendering them inherently observable-dependent.

%There had not been a thermodynamic relation among the normalized asymmetry of cross-correlations $\chi_{ab}$, dynamical activity $\mathcal{A}$ and entropy production rate $\sigma$, while binary relations between two ingredients have been reported.
Regarding  the dynamical activity and entropy production rate, both of them are greater than the pseudo entropy production rate \cite{shiraishi2023a}.
The reciprocal of the relative fluctuation  can be proved to be less than the pseudo entropy production \cite{shiraishi2021a,shiraishi2023a}, so the thermodynamic uncertainty relation and the kinetic uncertainty relation follow immediately.
This shows the ``duality" between the activity and entropy production rate, but not the relation between them.
The product of $\mathcal{A}$ and $\sigma$ appear in several studies.
For example, in the classical speed limit \cite{shiraishi2018}, $\mathcal{W}^2\le 2 \bar {\mathcal{A}} \sigma \tau$, where $\mathcal{W}$ is the Wasserstein distance between the initial and final probability distributions, $\bar{\mathcal{A}}$ the time-averaged activity, and $\tau$ the evolution time duration.
By contrast, the quotient of $\sigma$ over $\mathcal{A}$ appears in this study,  which arises naturally  from two perspectives.
From a dimensional analysis standpoint, $\chi_{ab}$ is dimensionless, while $\sigma$ has the dimension of rates. The dynamical activity $\mathcal{A}$ quantifies the inherent time-scale of the system with the dimension of rates too, so it seems intuitive to employ $\mathcal{A}$ as the denominator.
Furthermore, the numerator of the left-hand side, representing the asymmetry of cross-correlation, exhibits time-antisymmetry, while the denominator, corresponding to the geometric average of auto-correlation, displays time-symmetry.
As discussed earlier, the entropy production (dynamical activity) also exhibits (anti)symmetry with respect to time. 
By selecting the dynamical activity as the denominator, the structure of the normalized asymmetry is preserved.

As proved in Ref. \cite{ohga2023c}, the ratio between the real and imaginary components of the eigenvalue pertaining to the transition rate matrix can be regarded as a specific instance of the normalized asymmetry of cross-correlation.
Thus, the two bounds also provide insight into the spectra of transition rate matrix from a thermodynamic standpoint, in line with ongoing research along this direction \cite{barato2017,uhl2019,oberreiter2022,dechant2023,kolchinsky2024}  .
They could also contribute to the final resolution of the Oberreiter-Seifert-Barato conjecture \cite{oberreiter2022} as the two bounds directly incorporate the entropy production (in comparison to \cite{ohga2023c}) and are observable-independent (in comparison to \cite{shiraishi2023}).

%[Thermodynamic bounds on spectral perturbations]
%[Van 2023]

%[compare with Shiraishi and Van]

\begin{figure}[htb!]
  \includegraphics[width=\columnwidth]{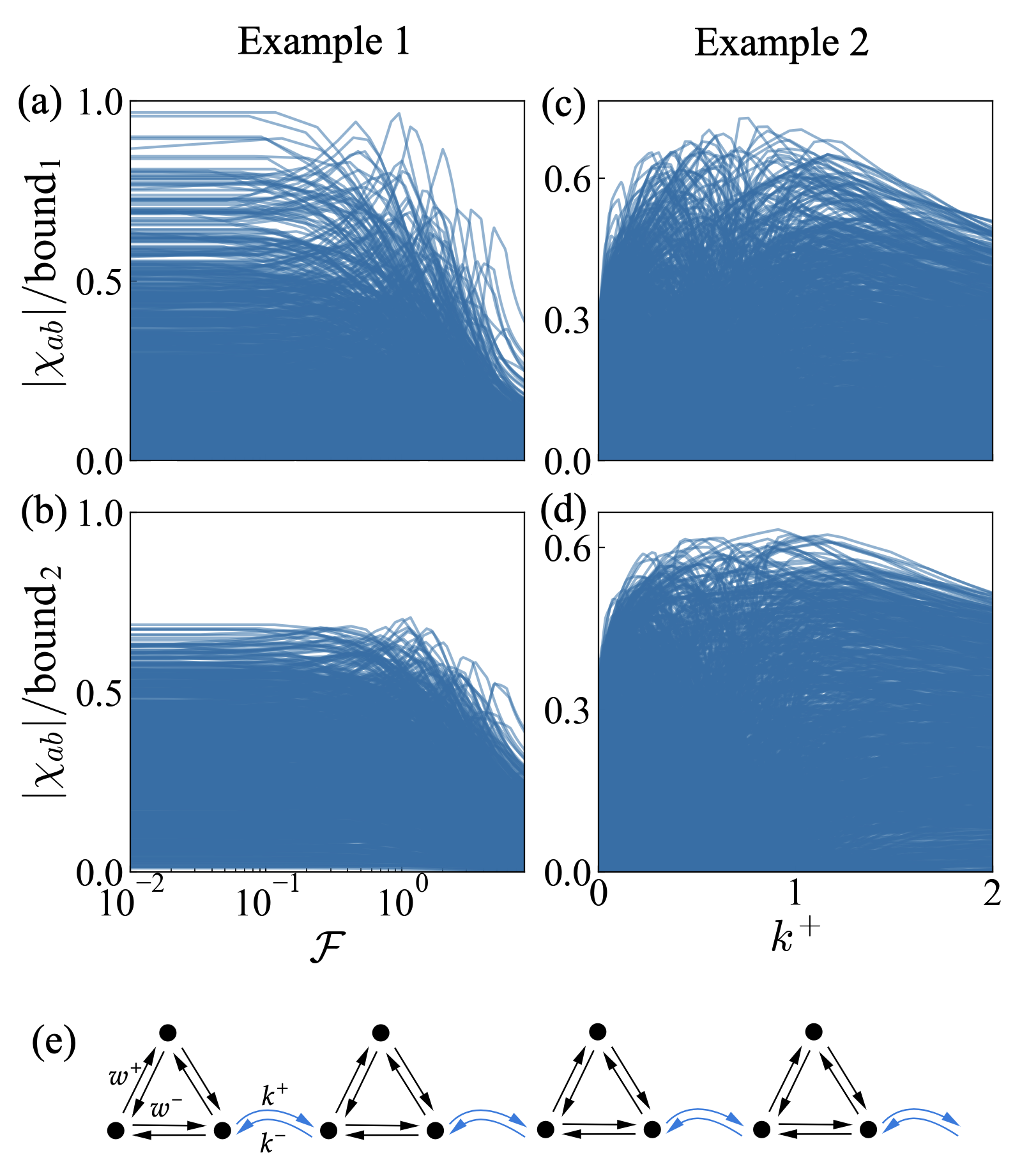}
 \caption{(a)--(d): Ratio between $|\chi_{ab}|$ and the bound as a function of the cycle affinity $\mathcal{F}$ (or $k^+$).
 The left (right) column corresponds to Example 1 (2), and the upper (lower) row corresponds to the first (second) bound.
(e) Schematic of the model in Example 2, where three states are grouped into a cell.
 }
 \label{fig:examples}
\end{figure}

\textit{Example 1.---}
As the first example, we consider the standard model of biochemical signal transduction as in Ref. \cite{ohga2023c}.
The system comprises an upstream receptor and a downstream protein. 
The upstream receptor undergoes stochastic switching between ``OFF" and ``ON" states, corresponding to the observable $a = 0, 1$. 
Similarly, the downstream protein stochastically switches between inactive and active states, corresponding to the observable $b = 0, 1$.
The dynamics of this  system is modeled by a four-state unicyclic Markov network [c.f. Fig. \ref{fig:scatter}(c)], whose transition rate matrix is given by
\begin{equation}
\begin{pmatrix}
		-k_b^{+,\text{OFF}}-k_a^+ & k_b^- & 0 & k_a^- \\
		k_b^{+,\text{OFF}} & -k_b^- - k_a^+  & k_a^- & 0 \\
		0 & k_a^+ & -k_a^- - k_b^- & k_b^{+,\text{ON}} \\
		k_a^+ & 0 & k_b^- & -k_a^- -k_b^{+,\text{ON}}
	\end{pmatrix}.
\end{equation}
According to stochastic thermodynamics \cite{seifert2012}, the cycle affinity is given by $\mathcal{F} = \ln (k_b^{+,\text{OFF}} /k_b^{+,\text{ON}})$.

We validate the bounds with Figs. \ref{fig:examples}(a) and (b).
Each curve in the figure is obtained as follows: $k_a^+, k_a^-, k_b^{+,\text{OFF}} $ and $k_b^-$ are sampled randomly from the uniform distribution, and the ratio of $|\chi_{ab}|$ to the bound is plotted versus the affinity $\mathcal{F}$ by varying $k_b^{+,\text{ON}}$.
This procedure is then repeated for $10^3$ times.
From Figs. \ref{fig:examples}(a) and (b), it can be seen that the two bounds are validated, and the first bound seems to be relatively tighter.
Cusps can be observed in nearly every curve. 
Their presence is not a result of the discretization process used for plotting. 
These cusps actually emerge due to the influence of the term $\gamma$ in the bounds:
Approaching a cusp point, a distinct combination of $\max \mathcal{A}_{mn}$ and $\min \mathcal{A}_{mn}$ takes over and alters the overall trend.
This is similar to how free energy changes at a first-order phase transition.
 By following the trend in one curve, it seems that without this mechanism, some ratios tend to surpass unity.

\textit{Example 2.---}
As an illustrative example  of multicyclic networks, we examine a
network  simplified from a  model of molecular motors  \cite{pietzonka2016a}.
The corresponding schematic is depicted in Fig. \ref{fig:examples}(e).
 In this network, three internal states are grouped into a ``cell," and the transitions occurring between adjacent cells signify either a forward or a backward step taken by the motor. 
 We consider a ring structure consisting of four such cells.
For simplicity,  we assume that the transition rates occurring both within the cell and between neighboring cells are uniform, as shown in the figure.
Following a similar approach as in Example 1, we generate each curve shown in Figs. \ref{fig:examples}(c) and (d) by sampling from the uniform distribution and subsequently fixing the values of $w^+$, $w^-$, and $k^-$. 
The quantity $|\chi_{ab}|$ is calculated as $|\lambda''/\lambda'|$, where $\lambda'$ $(\lambda'')$ is the real (imaginary) part of the eigenvalue (with the largest nonzero real part) of the transition matrix $\boldsymbol{W}$ \cite{ohga2023c}.
The plotted quantity corresponds to the ratio of $|\chi_{ab}|$ to its bound, with $k^+$ being systematically varied. 
This entire procedure is repeated a total of $10^3$ times.
The validity of the two bounds is demonstrated in Figs. \ref{fig:examples}(c) and (d), where behaviors similar to those observed in Example 1 are evident. As expected, the bound cannot be saturated for multicyclic networks.

\textit{Derivation.---}
For unicyclic networks $(1 \to 2 \to \ldots \to \mathcal{N} \to 1)$, 
The steady-state currents are uniform, i.e., $\mathcal{J}_{i,i+1} \equiv \mathcal{J}$.
Following the line of reasoning in Ref. \cite{ohga2023c}, it can be assumed that $a$ and $b$ are scaled to satisfy $D_a D_b = (D_a+D_b)^2/4$.
With this assumption, and employing Eqs. \eqref{eq:quantities}-\eqref{eq:slopes}, we arrive at
\begin{equation}
\label{eq:unicyclic}
	\begin{aligned}		\frac{\mathcal{A}\alpha_{ab}^2}{(D_a + D_b)^2/4} & =  \frac{16\sum_{i}\mathcal{A}_{i,i+1}(\sum_{j} \Omega_{j,j+1} )^2 \mathcal{J}^2}{(\sum_{i} \mathcal{A}_{i,i+1} L_{i,i+1}^{2})^2} \\
&\le {16\gamma^2}\left(\sum_{i} \frac{\mathcal{J}^2}{\mathcal{A}_{i,i+1}} \right) \frac{(\sum_{i} \Omega_{i,i+1} )^2 }{( \sum_iL_{i,i+1}^{2})^2} \\
& \le \frac{\gamma^2 \sigma}{2\tan^2 (\pi/\mathcal{N})},  
 	\end{aligned}
\end{equation}
which is equivalent to Eq. \eqref{eq:main}. 
Note that for this case $\gamma = \max \mathcal{A}_{i,i+1}/\min \mathcal{A}_{i,i+1}$ as defined in Eq. \eqref{eq:gamma}.
In the derivation of the first inequality, we used the inequality
\begin{equation}
\frac{\sum x_i}{(\sum x_i y_i)^2} 
\le \frac{\sum [(\max \boldsymbol{x})^2/x_i]}{(\min \boldsymbol{x}\sum  y_i)^2} 
\le \left (\frac{\max \boldsymbol{x}}{\min \boldsymbol{x}}\right)^2 \frac{\sum x_i^{-1}}{(\sum y_i)^2}	,
\end{equation}
  where $x_i,y_i >0$. 
  The equality  holds if the local activities are uniform, i.e., $\mathcal{A}_{12} = \mathcal{A}_{23} = \ldots = \mathcal{A}_{\mathcal{N}1}$.
A sufficient condition for this is uniform transition rates, i.e., $W_{12}=W_{23}=\ldots=W_{\mathcal{N}1}$ and $W_{21}=W_{32}=\ldots=W_{1\mathcal{N}}$.
The term $2\sum_{i} {\mathcal{J}^2}/{\mathcal{A}_{i,i+1}}$ can be identified as the pseudo entropy production rate \cite{shiraishi2018,shiraishi2021a,shiraishi2023a}, which is always equal to or less than the actual entropy production rate.
They coincide when both vanish.
With the pseudo entropy production rate, to obtain the last inequality, we first used the Cauchy-Schwarz inequality $\sum_iL_{i,i+1}^{2} \ge  (\sum_i L_{i,i+1})^2/{\mathcal{N}}$,  followed by the isoperimetric inequality
  \begin{equation}
  	\left(4 \mathcal{N} \tan \frac{\pi}{\mathcal{N}}\right)\bigg|\sum_{i} \Omega_{i,i+1}\bigg| \leq\bigg(\sum_{i} L_{i,i+1}\bigg)^{2}.
  \end{equation}
The equalities in Cauchy-Schwarz and isoperimetric inequality  hold simultaneously if and only if the points $(a_i,b_i)$ form a regular polygon.

For bound \eqref{eq:main2}, we begin with Eq. (S31) in Ref. \cite{ohga2023c}:
\begin{equation}
\left|\chi_{b a}\right| \leq \frac{4 \sum_{c \in \mathcal{C}^{*}} \mathcal{J}_{c}\left(\sum_{e \in c} L_{e}\right)^{2}\left[4 n_{c}^{\prime} \tan \left(\pi / n_{c}^{\prime}\right)\right]^{-1}}{\sum_{c \in \mathcal{C}^{*}} \mathcal{J}_{c}\left(\sum_{e \in c} L_{e}\right)^{2}\left[n_{c} \tanh \left(\mathcal{F}_{c} / 2 n_{c}\right)\right]^{-1}}  .
\end{equation}
Here, $\mathcal{C}^*$ is the set of cycles  with nonzero net asymmetry generated by the uniform cycle decomposition \cite{pietzonka2016}.
Given a cycle $c$,
 $n_c$ is the number of states, $n_c'$ is the number of times the joint value $(a, b)$ changes throughout the duration of the cycle, $\mathcal{J}_c$ is the cycle current, and $\mathcal{F}_c$ is the cycle affinity.
For more details of $n_c'$, $n_c$ and $\mathcal{C}^*$, please refer to Ref. \cite{ohga2023c}.
Since $\left[4 n_{c}^{\prime} \tan \left(\pi / n_{c}^{\prime}\right)\right]^{-1}$ is a monotonically increasing function of $n_c'$ $(n_c'>2)$ and $\left[n_{c} \tanh \left(\mathcal{F}_{c} / 2 n_{c}\right)\right]^{-1}$ a decreasing function of $n_c$ $(n_c >0)$, we have
\begin{equation}
\begin{aligned}
\left|\chi_{b a}\right| & \le \frac{4 \sum_{c \in \mathcal{C}^{*}} \mathcal{J}_{c}\left(\sum_{e \in c} L_{e}\right)^{2}  \left[\mathcal{N} \tanh \left(\mathcal{F}_{c} / 2 \mathcal{N}\right)\right]}{\sum_{c \in \mathcal{C}^{*}} \mathcal{J}_{c}\left(\sum_{e \in c} L_{e}\right)^{2}  \left[4 \mathcal{N} \tan \left(\pi / \mathcal{N}\right)\right]} 
\\
& \le   \frac{ \sum_{c \in \mathcal{C}^{*}} \sigma_c\left(\sum_{e \in c} L_{e}\right)^{2}}{2\pi\sum_{c \in \mathcal{C}^{*}} \mathcal{J}_{c}\left(\sum_{e \in c} L_{e}\right)^{2}} ,
\end{aligned}
\label{eq:chi1}
\end{equation}
where we have used $n_c' \le n_c \le \mathcal{N}$, $\mathcal{N} \tanh (\mathcal{F}_c/2\mathcal{N}) /[4\mathcal{N} \tan (\pi/\mathcal{N})] \le \mathcal{F}_c /8\pi$  and $\sigma_c = \mathcal{F}_c \mathcal{J}_c$.
The equality holds for unicyclic systems when the observables form a regular polygon in the large $\mathcal{N}$ limit.

 On the other hand, one can use Eq. (S10) in Ref. \cite{ohga2023c} and the fact that $\mathcal{C}^*$ is a restricted set to bound  the normalized asymmetry by
 \begin{equation}
 \left|\chi_{b a}\right| \le \frac{4\sum_{c \in \mathcal{C}^{*}} \mathcal{J}_{c}\left|\sum_{e \in c} \Omega_{e}\right|}{  \sum_{c \in \mathcal{C}^{*}}\sum_{e \in c} \mathcal{A}_e L_{e}^2}.
 \end{equation}
 Furthermore, we obtain
\begin{equation}
\begin{aligned}
\left|\chi_{b a}\right|&  \leq \frac{4\sum_{c \in \mathcal{C}^{*}} \mathcal{J}_{c}\left(\sum_{e \in c} L_{e}\right)^{2}\left[4 n_{c}^{\prime} \tan \left(\pi / n_{c}^{\prime}\right)\right]^{-1}}{\min \mathcal{A}_{mn}\sum_{c \in \mathcal{C}^{*}} \sum_{e \in c} L_{e}^{2}} \\
 & \leq \frac{\sum_{c \in \mathcal{C}^{*}} \mathcal{J}_{c}\left(\sum_{e \in c} L_{e}\right)^{2}}{\tan(\pi/\mathcal{N})\min \mathcal{A}_{mn}\sum_{c \in \mathcal{C}^{*}} \left(\sum_{e \in c} L_{e}\right)^{2}}  ,
\end{aligned}
\label{eq:chi2}
\end{equation}
where we have used the isoperimetric inequality, the monotonicity of $\left[4 n_{c}^{\prime} \tan \left(\pi / n_{c}^{\prime}\right)\right]^{-1}$  and Cauchy-Schwarz inequality again.
The equality holds for unicyclic systems when the observables form a regular polygon.

 Multiplying these two bounds on $|\chi_{ab}|$ [Eqs. \eqref{eq:chi1} and \eqref{eq:chi2}] and employing $\mathcal{A} \leq \mathcal{N}^* \max \mathcal{A}_{mn}$  results in
\begin{equation}
\begin{aligned}
\chi_{b a}^2 & \le \frac{\sum_{c \in \mathcal{C}^{*}} \sigma_c\left(\sum_{e \in c} L_{e}\right)^{2}}{2\pi\tan(\pi/\mathcal{N})\min \mathcal{A}_{mn}\sum_{c \in C^{*}} \left(\sum_{e \in c} L_{e}\right)^{2}} \\
&\le \frac{\mathcal{N}^* \gamma }{2\pi\tan (\pi/\mathcal{N})} \frac{\sigma}{\mathcal{A}},
\end{aligned}
\end{equation}
which is equivalent to Eq. \eqref{eq:main2}.
Here, $\mathcal{N}^*$ is the number of nonzero $\mathcal{A}_{mn}$'s.
By definition $\mathcal{A}_{mn} = \mathcal{A}_{nm}$, so each pair is counted as one.
For unicyclic networks, $\mathcal{N}^* =\mathcal{N} $.

%Considering the connection between the affinity and entropy production rate, one may wonder whether our result is implied by Eq. \eqref{eq:ohga}.
%For uniform unicyclic networks, we have
%\begin{equation}
%\label{eq:derived}
%	\begin{aligned}
%		\frac{\alpha_{ab}^2}{{(D_a + D_b)^2/4}} &\le \frac{2\mathcal{F }\mathcal{J} |\sum_{i} \Omega_{i,i+1}|}{\pi\sum_i L_{i,i+1}^2 \mathcal{A}_{i,i+1}} \\
%		& \le \frac{2\mathcal{N} \gamma \sigma}{\pi \mathcal{A}}  \frac{|\sum_{i} \Omega_{i,i+1} | }{ \sum_iL_{i,i+1}^{2}} \\
%	 &	\le \frac{\mathcal{N}\gamma^2}{2\pi \tan(\pi/\mathcal{N})}   \frac{\sigma}{\mathcal{A}}.
%	\end{aligned}
%\end{equation}
%The looser bound in Eq. \eqref{eq:ohga} has been used to obtain the first inequality. 
%For the second and last inequality,  we have used $\sum \mathcal{A}_{i,i+1} \leq \mathcal{N} \max \mathcal{A}_{i,i+1}$ and $\gamma \ge 1$, along with manipulations that were used to derive Eq. \eqref{eq:unicyclic}.
%The inequality above has the same form with Eq. \eqref{eq:unicyclic} except for the prefactor.
%Since $\tan x>x$ for $x \in [0, \pi/2)$, Eq. \eqref{eq:derived} is always looser than Eq. \eqref{eq:unicyclic}, while they are asymptotically equivalent to each other in the large $\mathcal{N}$ limit.
%Of course this does not exclude the possibility to  of deriving Eq. \eqref{eq:main} from Eq. \eqref{eq:ohga}, but it does not seem straightforward even for unicyclic networks.
% \newpage

\textit{Discussion.---}
In summary, we report two thermodynamic bounds [Eqs. \eqref{eq:main} and \eqref{eq:main2}] on the normalized asymmetry of cross-correlation in terms of the dynamical activity and entropy production.
Identifying a simpler expression that directly relates these three quantities seems to be a challenge.
This bound exhibits broad applicability, regardless of the underlying network topology, and offers experimental testability.
We also anticipate that this bound will  provide valuable insights in addressing the conjecture proposed by Oberreiter \textit{et al.} \cite{oberreiter2022}.
Rigorously establishing the validity of Eq. \eqref{eq:main} for arbitrary network topologies does not appear to be straightforward, but to this end, the utilization of uniform cycle decomposition may serve as a crucial technique \cite{pietzonka2016,ohga2023c}.  
In view of the Langevin equation being regarded as the continuous-space limit of the master equation, we anticipate that our two bounds hold for overdamped Langevin systems, while additional terms might be needed for underdamped cases \cite{vanvu2019,fu2022}.  
A future research direction entails investigating whether this bound holds true or how it should be modified when applied to partially observed Markov networks \cite{vandermeer2022,ghosal2022,harunari2022}. 
Additionally, exploring the influence of quantum effects, such as quantum coherence,  on this relation represents an intriguing avenue for further investigation.

\emph{Acknowledgment.---}
We thank Kangqiao Liu for valuable discussions. 

%\bibliography{ref}

\begin{thebibliography}{53}%
\makeatletter
\providecommand \@ifxundefined [1]{%
 \@ifx{#1\undefined}
}%
\providecommand \@ifnum [1]{%
 \ifnum #1\expandafter \@firstoftwo
 \else \expandafter \@secondoftwo
 \fi
}%
\providecommand \@ifx [1]{%
 \ifx #1\expandafter \@firstoftwo
 \else \expandafter \@secondoftwo
 \fi
}%
\providecommand \natexlab [1]{#1}%
\providecommand \enquote  [1]{``#1''}%
\providecommand \bibnamefont  [1]{#1}%
\providecommand \bibfnamefont [1]{#1}%
\providecommand \citenamefont [1]{#1}%
\providecommand \href@noop [0]{\@secondoftwo}%
\providecommand \href [0]{\begingroup \@sanitize@url \@href}%
\providecommand \@href[1]{\@@startlink{#1}\@@href}%
\providecommand \@@href[1]{\endgroup#1\@@endlink}%
\providecommand \@sanitize@url [0]{\catcode `\\12\catcode `\$12\catcode
  `\&12\catcode `\#12\catcode `\^12\catcode `\_12\catcode `\%12\relax}%
\providecommand \@@startlink[1]{}%
\providecommand \@@endlink[0]{}%
\providecommand \url  [0]{\begingroup\@sanitize@url \@url }%
\providecommand \@url [1]{\endgroup\@href {#1}{\urlprefix }}%
\providecommand \urlprefix  [0]{URL }%
\providecommand \Eprint [0]{\href }%
\providecommand \doibase [0]{https://doi.org/}%
\providecommand \selectlanguage [0]{\@gobble}%
\providecommand \bibinfo  [0]{\@secondoftwo}%
\providecommand \bibfield  [0]{\@secondoftwo}%
\providecommand \translation [1]{[#1]}%
\providecommand \BibitemOpen [0]{}%
\providecommand \bibitemStop [0]{}%
\providecommand \bibitemNoStop [0]{.\EOS\space}%
\providecommand \EOS [0]{\spacefactor3000\relax}%
\providecommand \BibitemShut  [1]{\csname bibitem#1\endcsname}%
\let\auto@bib@innerbib\@empty
%</preamble>
\bibitem [{\citenamefont {Landi}\ and\ \citenamefont
  {Paternostro}(2021)}]{landi2021}%
  \BibitemOpen
  \bibfield  {author} {\bibinfo {author} {\bibfnamefont {G.~T.}\ \bibnamefont
  {Landi}}\ and\ \bibinfo {author} {\bibfnamefont {M.}~\bibnamefont
  {Paternostro}},\ }\bibfield  {title} {\bibinfo {title} {Irreversible entropy
  production, from quantum to classical},\ }\href
  {https://doi.org/10.1103/RevModPhys.93.035008} {\bibfield  {journal}
  {\bibinfo  {journal} {Rev. Mod. Phys.}\ }\textbf {\bibinfo {volume} {93}},\
  \bibinfo {pages} {035008} (\bibinfo {year} {2021})}\BibitemShut {NoStop}%
\bibitem [{\citenamefont {Ruppeiner}(1995)}]{ruppeiner1995}%
  \BibitemOpen
  \bibfield  {author} {\bibinfo {author} {\bibfnamefont {G.}~\bibnamefont
  {Ruppeiner}},\ }\bibfield  {title} {\bibinfo {title} {Riemannian geometry in
  thermodynamic fluctuation theory},\ }\href
  {https://doi.org/10.1103/RevModPhys.67.605} {\bibfield  {journal} {\bibinfo
  {journal} {Rev. Mod. Phys.}\ }\textbf {\bibinfo {volume} {67}},\ \bibinfo
  {pages} {605} (\bibinfo {year} {1995})}\BibitemShut {NoStop}%
\bibitem [{\citenamefont {Crooks}(2007)}]{crooks2007}%
  \BibitemOpen
  \bibfield  {author} {\bibinfo {author} {\bibfnamefont {G.~E.}\ \bibnamefont
  {Crooks}},\ }\bibfield  {title} {\bibinfo {title} {Measuring {{Thermodynamic
  Length}}},\ }\href {https://doi.org/10.1103/PhysRevLett.99.100602} {\bibfield
   {journal} {\bibinfo  {journal} {Phys. Rev. Lett.}\ }\textbf {\bibinfo
  {volume} {99}},\ \bibinfo {pages} {100602} (\bibinfo {year}
  {2007})}\BibitemShut {NoStop}%
\bibitem [{\citenamefont {Sivak}\ and\ \citenamefont
  {Crooks}(2012)}]{sivak2012}%
  \BibitemOpen
  \bibfield  {author} {\bibinfo {author} {\bibfnamefont {D.~A.}\ \bibnamefont
  {Sivak}}\ and\ \bibinfo {author} {\bibfnamefont {G.~E.}\ \bibnamefont
  {Crooks}},\ }\bibfield  {title} {\bibinfo {title} {Thermodynamic {{Metrics}}
  and {{Optimal Paths}}},\ }\href
  {https://doi.org/10.1103/PhysRevLett.108.190602} {\bibfield  {journal}
  {\bibinfo  {journal} {Phys. Rev. Lett.}\ }\textbf {\bibinfo {volume} {108}},\
  \bibinfo {pages} {190602} (\bibinfo {year} {2012})}\BibitemShut {NoStop}%
\bibitem [{\citenamefont {Gu}(2023{\natexlab{a}})}]{gu2023}%
  \BibitemOpen
  \bibfield  {author} {\bibinfo {author} {\bibfnamefont {J.}~\bibnamefont
  {Gu}},\ }\bibfield  {title} {\bibinfo {title} {Work statistics in slow
  thermodynamic processes},\ }\href {https://doi.org/10.1063/5.0138405}
  {\bibfield  {journal} {\bibinfo  {journal} {J. Chem. Phys.}\ }\textbf
  {\bibinfo {volume} {158}},\ \bibinfo {pages} {074104} (\bibinfo {year}
  {2023}{\natexlab{a}})}\BibitemShut {NoStop}%
\bibitem [{\citenamefont {Shiraishi}\ \emph {et~al.}(2018)\citenamefont
  {Shiraishi}, \citenamefont {Funo},\ and\ \citenamefont
  {Saito}}]{shiraishi2018}%
  \BibitemOpen
  \bibfield  {author} {\bibinfo {author} {\bibfnamefont {N.}~\bibnamefont
  {Shiraishi}}, \bibinfo {author} {\bibfnamefont {K.}~\bibnamefont {Funo}},\
  and\ \bibinfo {author} {\bibfnamefont {K.}~\bibnamefont {Saito}},\ }\bibfield
   {title} {\bibinfo {title} {Speed {{Limit}} for {{Classical Stochastic
  Processes}}},\ }\href {https://doi.org/10.1103/PhysRevLett.121.070601}
  {\bibfield  {journal} {\bibinfo  {journal} {Phys. Rev. Lett.}\ }\textbf
  {\bibinfo {volume} {121}},\ \bibinfo {pages} {070601} (\bibinfo {year}
  {2018})}\BibitemShut {NoStop}%
\bibitem [{\citenamefont {Vo}\ \emph {et~al.}(2020)\citenamefont {Vo},
  \citenamefont {Van~Vu},\ and\ \citenamefont {Hasegawa}}]{vo2020}%
  \BibitemOpen
  \bibfield  {author} {\bibinfo {author} {\bibfnamefont {V.~T.}\ \bibnamefont
  {Vo}}, \bibinfo {author} {\bibfnamefont {T.}~\bibnamefont {Van~Vu}},\ and\
  \bibinfo {author} {\bibfnamefont {Y.}~\bibnamefont {Hasegawa}},\ }\bibfield
  {title} {\bibinfo {title} {Unified {{Approach}} to {{Classical Speed Limit}}
  and {{Thermodynamic Uncertainty Relation}}},\ }\href
  {https://doi.org/10.1103/PhysRevE.102.062132} {\bibfield  {journal} {\bibinfo
   {journal} {Phys. Rev. E}\ }\textbf {\bibinfo {volume} {102}},\ \bibinfo
  {pages} {062132} (\bibinfo {year} {2020})}\BibitemShut {NoStop}%
\bibitem [{\citenamefont {Dechant}(2022)}]{dechant2022a}%
  \BibitemOpen
  \bibfield  {author} {\bibinfo {author} {\bibfnamefont {A.}~\bibnamefont
  {Dechant}},\ }\bibfield  {title} {\bibinfo {title} {Minimum entropy
  production, detailed balance and {{Wasserstein}} distance for continuous-time
  {{Markov}} processes},\ }\href {https://doi.org/10.1088/1751-8121/ac4ac0}
  {\bibfield  {journal} {\bibinfo  {journal} {J. Phys. A: Math. Theor.}\
  }\textbf {\bibinfo {volume} {55}},\ \bibinfo {pages} {094001} (\bibinfo
  {year} {2022})}\BibitemShut {NoStop}%
\bibitem [{\citenamefont {Nicholson}\ \emph {et~al.}(2020)\citenamefont
  {Nicholson}, \citenamefont {{Garc{\'i}a-Pintos}}, \citenamefont {{del
  Campo}},\ and\ \citenamefont {Green}}]{nicholson2020}%
  \BibitemOpen
  \bibfield  {author} {\bibinfo {author} {\bibfnamefont {S.~B.}\ \bibnamefont
  {Nicholson}}, \bibinfo {author} {\bibfnamefont {L.~P.}\ \bibnamefont
  {{Garc{\'i}a-Pintos}}}, \bibinfo {author} {\bibfnamefont {A.}~\bibnamefont
  {{del Campo}}},\ and\ \bibinfo {author} {\bibfnamefont {J.~R.}\ \bibnamefont
  {Green}},\ }\bibfield  {title} {\bibinfo {title} {Time\textendash information
  uncertainty relations in thermodynamics},\ }\href
  {https://doi.org/10.1038/s41567-020-0981-y} {\bibfield  {journal} {\bibinfo
  {journal} {Nat. Phys.}\ }\textbf {\bibinfo {volume} {16}},\ \bibinfo {pages}
  {1211} (\bibinfo {year} {2020})}\BibitemShut {NoStop}%
\bibitem [{\citenamefont {{Garc{\'i}a-Pintos}}\ \emph
  {et~al.}(2022)\citenamefont {{Garc{\'i}a-Pintos}}, \citenamefont {Nicholson},
  \citenamefont {Green}, \citenamefont {{del Campo}},\ and\ \citenamefont
  {Gorshkov}}]{garcia-pintos2022}%
  \BibitemOpen
  \bibfield  {author} {\bibinfo {author} {\bibfnamefont {L.~P.}\ \bibnamefont
  {{Garc{\'i}a-Pintos}}}, \bibinfo {author} {\bibfnamefont {S.~B.}\
  \bibnamefont {Nicholson}}, \bibinfo {author} {\bibfnamefont {J.~R.}\
  \bibnamefont {Green}}, \bibinfo {author} {\bibfnamefont {A.}~\bibnamefont
  {{del Campo}}},\ and\ \bibinfo {author} {\bibfnamefont {A.~V.}\ \bibnamefont
  {Gorshkov}},\ }\bibfield  {title} {\bibinfo {title} {Unifying {{Quantum}} and
  {{Classical Speed Limits}} on {{Observables}}},\ }\href
  {https://doi.org/10.1103/PhysRevX.12.011038} {\bibfield  {journal} {\bibinfo
  {journal} {Phys. Rev. X}\ }\textbf {\bibinfo {volume} {12}},\ \bibinfo
  {pages} {011038} (\bibinfo {year} {2022})}\BibitemShut {NoStop}%
\bibitem [{\citenamefont {Gu}(2023{\natexlab{b}})}]{gu2023a}%
  \BibitemOpen
  \bibfield  {author} {\bibinfo {author} {\bibfnamefont {J.}~\bibnamefont
  {Gu}},\ }\bibfield  {title} {\bibinfo {title} {Speed limit, dissipation
  bound, and dissipation-time trade-off in thermal relaxation processes},\
  }\href {https://doi.org/10.1103/PhysRevE.108.L052103} {\bibfield  {journal}
  {\bibinfo  {journal} {Phys. Rev. E}\ }\textbf {\bibinfo {volume} {108}},\
  \bibinfo {pages} {L052103} (\bibinfo {year}
  {2023}{\natexlab{b}})}\BibitemShut {NoStop}%
\bibitem [{\citenamefont {Barato}\ and\ \citenamefont
  {Seifert}(2015)}]{barato2015}%
  \BibitemOpen
  \bibfield  {author} {\bibinfo {author} {\bibfnamefont {A.~C.}\ \bibnamefont
  {Barato}}\ and\ \bibinfo {author} {\bibfnamefont {U.}~\bibnamefont
  {Seifert}},\ }\bibfield  {title} {\bibinfo {title} {Thermodynamic
  {{Uncertainty Relation}} for {{Biomolecular Processes}}},\ }\href
  {https://doi.org/10.1103/PhysRevLett.114.158101} {\bibfield  {journal}
  {\bibinfo  {journal} {Phys. Rev. Lett.}\ }\textbf {\bibinfo {volume} {114}},\
  \bibinfo {pages} {158101} (\bibinfo {year} {2015})}\BibitemShut {NoStop}%
\bibitem [{\citenamefont {Gingrich}\ \emph {et~al.}(2016)\citenamefont
  {Gingrich}, \citenamefont {Horowitz}, \citenamefont {Perunov},\ and\
  \citenamefont {England}}]{gingrich2016}%
  \BibitemOpen
  \bibfield  {author} {\bibinfo {author} {\bibfnamefont {T.~R.}\ \bibnamefont
  {Gingrich}}, \bibinfo {author} {\bibfnamefont {J.~M.}\ \bibnamefont
  {Horowitz}}, \bibinfo {author} {\bibfnamefont {N.}~\bibnamefont {Perunov}},\
  and\ \bibinfo {author} {\bibfnamefont {J.~L.}\ \bibnamefont {England}},\
  }\bibfield  {title} {\bibinfo {title} {Dissipation {{Bounds All Steady-State
  Current Fluctuations}}},\ }\href
  {https://doi.org/10.1103/PhysRevLett.116.120601} {\bibfield  {journal}
  {\bibinfo  {journal} {Phys. Rev. Lett.}\ }\textbf {\bibinfo {volume} {116}},\
  \bibinfo {pages} {120601} (\bibinfo {year} {2016})}\BibitemShut {NoStop}%
\bibitem [{\citenamefont {Horowitz}\ and\ \citenamefont
  {Gingrich}(2020)}]{horowitz2020}%
  \BibitemOpen
  \bibfield  {author} {\bibinfo {author} {\bibfnamefont {J.~M.}\ \bibnamefont
  {Horowitz}}\ and\ \bibinfo {author} {\bibfnamefont {T.~R.}\ \bibnamefont
  {Gingrich}},\ }\bibfield  {title} {\bibinfo {title} {Thermodynamic
  uncertainty relations constrain non-equilibrium fluctuations},\ }\href
  {https://doi.org/10.1038/s41567-019-0702-6} {\bibfield  {journal} {\bibinfo
  {journal} {Nat. Phys.}\ }\textbf {\bibinfo {volume} {16}},\ \bibinfo {pages}
  {15} (\bibinfo {year} {2020})}\BibitemShut {NoStop}%
\bibitem [{\citenamefont {Tomita}\ and\ \citenamefont
  {Tomita}(1974)}]{tomita1974a}%
  \BibitemOpen
  \bibfield  {author} {\bibinfo {author} {\bibfnamefont {K.}~\bibnamefont
  {Tomita}}\ and\ \bibinfo {author} {\bibfnamefont {H.}~\bibnamefont
  {Tomita}},\ }\bibfield  {title} {\bibinfo {title} {Irreversible
  {{Circulation}} of {{Fluctuation}}},\ }\href
  {https://doi.org/10.1143/PTP.51.1731} {\bibfield  {journal} {\bibinfo
  {journal} {Progress of Theoretical Physics}\ }\textbf {\bibinfo {volume}
  {51}},\ \bibinfo {pages} {1731} (\bibinfo {year} {1974})}\BibitemShut
  {NoStop}%
\bibitem [{\citenamefont {Tomita}\ and\ \citenamefont
  {Sano}(2008)}]{tomita2008}%
  \BibitemOpen
  \bibfield  {author} {\bibinfo {author} {\bibfnamefont {H.}~\bibnamefont
  {Tomita}}\ and\ \bibinfo {author} {\bibfnamefont {M.~M.}\ \bibnamefont
  {Sano}},\ }\bibfield  {title} {\bibinfo {title} {Irreversible {{Circulation}}
  of {{Fluctuation}} and {{Entropy Production}}},\ }\href
  {https://doi.org/10.1143/PTP.119.515} {\bibfield  {journal} {\bibinfo
  {journal} {Progress of Theoretical Physics}\ }\textbf {\bibinfo {volume}
  {119}},\ \bibinfo {pages} {515} (\bibinfo {year} {2008})}\BibitemShut
  {NoStop}%
\bibitem [{\citenamefont {Ohga}\ \emph {et~al.}(2023)\citenamefont {Ohga},
  \citenamefont {Ito},\ and\ \citenamefont {Kolchinsky}}]{ohga2023c}%
  \BibitemOpen
  \bibfield  {author} {\bibinfo {author} {\bibfnamefont {N.}~\bibnamefont
  {Ohga}}, \bibinfo {author} {\bibfnamefont {S.}~\bibnamefont {Ito}},\ and\
  \bibinfo {author} {\bibfnamefont {A.}~\bibnamefont {Kolchinsky}},\ }\bibfield
   {title} {\bibinfo {title} {Thermodynamic {{Bound}} on the {{Asymmetry}} of
  {{Cross-Correlations}}},\ }\href
  {https://doi.org/10.1103/PhysRevLett.131.077101} {\bibfield  {journal}
  {\bibinfo  {journal} {Phys. Rev. Lett.}\ }\textbf {\bibinfo {volume} {131}},\
  \bibinfo {pages} {077101} (\bibinfo {year} {2023})}\BibitemShut {NoStop}%
\bibitem [{\citenamefont {Barato}\ and\ \citenamefont
  {Seifert}(2017)}]{barato2017}%
  \BibitemOpen
  \bibfield  {author} {\bibinfo {author} {\bibfnamefont {A.~C.}\ \bibnamefont
  {Barato}}\ and\ \bibinfo {author} {\bibfnamefont {U.}~\bibnamefont
  {Seifert}},\ }\bibfield  {title} {\bibinfo {title} {Coherence of biochemical
  oscillations is bounded by driving force and network topology},\ }\href
  {https://doi.org/10.1103/PhysRevE.95.062409} {\bibfield  {journal} {\bibinfo
  {journal} {Phys. Rev. E}\ }\textbf {\bibinfo {volume} {95}},\ \bibinfo
  {pages} {062409} (\bibinfo {year} {2017})}\BibitemShut {NoStop}%
\bibitem [{\citenamefont {Oberreiter}\ \emph {et~al.}(2022)\citenamefont
  {Oberreiter}, \citenamefont {Seifert},\ and\ \citenamefont
  {Barato}}]{oberreiter2022}%
  \BibitemOpen
  \bibfield  {author} {\bibinfo {author} {\bibfnamefont {L.}~\bibnamefont
  {Oberreiter}}, \bibinfo {author} {\bibfnamefont {U.}~\bibnamefont
  {Seifert}},\ and\ \bibinfo {author} {\bibfnamefont {A.~C.}\ \bibnamefont
  {Barato}},\ }\bibfield  {title} {\bibinfo {title} {Universal minimal cost of
  coherent biochemical oscillations},\ }\href
  {https://doi.org/10.1103/PhysRevE.106.014106} {\bibfield  {journal} {\bibinfo
   {journal} {Phys. Rev. E}\ }\textbf {\bibinfo {volume} {106}},\ \bibinfo
  {pages} {014106} (\bibinfo {year} {2022})}\BibitemShut {NoStop}%
\bibitem [{\citenamefont {Shiraishi}(2023{\natexlab{a}})}]{shiraishi2023}%
  \BibitemOpen
  \bibfield  {author} {\bibinfo {author} {\bibfnamefont {N.}~\bibnamefont
  {Shiraishi}},\ }\bibfield  {title} {\bibinfo {title} {Entropy production
  limits all fluctuation oscillations},\ }\href
  {https://doi.org/10.1103/PhysRevE.108.L042103} {\bibfield  {journal}
  {\bibinfo  {journal} {Phys. Rev. E}\ }\textbf {\bibinfo {volume} {108}},\
  \bibinfo {pages} {L042103} (\bibinfo {year}
  {2023}{\natexlab{a}})}\BibitemShut {NoStop}%
\bibitem [{\citenamefont {Liang}\ and\ \citenamefont
  {Pigolotti}(2023)}]{liang2023}%
  \BibitemOpen
  \bibfield  {author} {\bibinfo {author} {\bibfnamefont {S.}~\bibnamefont
  {Liang}}\ and\ \bibinfo {author} {\bibfnamefont {S.}~\bibnamefont
  {Pigolotti}},\ }\bibfield  {title} {\bibinfo {title} {Thermodynamic bounds on
  time-reversal asymmetry},\ }\href
  {https://doi.org/10.1103/PhysRevE.108.L062101} {\bibfield  {journal}
  {\bibinfo  {journal} {Phys. Rev. E}\ }\textbf {\bibinfo {volume} {108}},\
  \bibinfo {pages} {L062101} (\bibinfo {year} {2023})}\BibitemShut {NoStop}%
\bibitem [{\citenamefont {Vu}\ \emph {et~al.}(2023)\citenamefont {Vu},
  \citenamefont {Vo},\ and\ \citenamefont {Saito}}]{vanvu2023b}%
  \BibitemOpen
  \bibfield  {author} {\bibinfo {author} {\bibfnamefont {T.~V.}\ \bibnamefont
  {Vu}}, \bibinfo {author} {\bibfnamefont {V.~T.}\ \bibnamefont {Vo}},\ and\
  \bibinfo {author} {\bibfnamefont {K.}~\bibnamefont {Saito}},\ }\href@noop {}
  {\bibinfo {title} {Dissipation, quantum coherence, and asymmetry of
  finite-time cross-correlations}} (\bibinfo {year} {2023}),\ \Eprint
  {https://arxiv.org/abs/2305.18000} {arXiv:2305.18000} \BibitemShut {NoStop}%
\bibitem [{\citenamefont {Maes}\ and\ \citenamefont
  {Van~Wieren}(2006)}]{maes2006}%
  \BibitemOpen
  \bibfield  {author} {\bibinfo {author} {\bibfnamefont {C.}~\bibnamefont
  {Maes}}\ and\ \bibinfo {author} {\bibfnamefont {M.~H.}\ \bibnamefont
  {Van~Wieren}},\ }\bibfield  {title} {\bibinfo {title} {Time-{{Symmetric
  Fluctuations}} in {{Nonequilibrium Systems}}},\ }\href
  {https://doi.org/10.1103/PhysRevLett.96.240601} {\bibfield  {journal}
  {\bibinfo  {journal} {Phys. Rev. Lett.}\ }\textbf {\bibinfo {volume} {96}},\
  \bibinfo {pages} {240601} (\bibinfo {year} {2006})}\BibitemShut {NoStop}%
\bibitem [{\citenamefont {Baiesi}\ \emph {et~al.}(2009)\citenamefont {Baiesi},
  \citenamefont {Maes},\ and\ \citenamefont {Wynants}}]{baiesi2009a}%
  \BibitemOpen
  \bibfield  {author} {\bibinfo {author} {\bibfnamefont {M.}~\bibnamefont
  {Baiesi}}, \bibinfo {author} {\bibfnamefont {C.}~\bibnamefont {Maes}},\ and\
  \bibinfo {author} {\bibfnamefont {B.}~\bibnamefont {Wynants}},\ }\bibfield
  {title} {\bibinfo {title} {Fluctuations and {{Response}} of {{Nonequilibrium
  States}}},\ }\href {https://doi.org/10.1103/PhysRevLett.103.010602}
  {\bibfield  {journal} {\bibinfo  {journal} {Phys. Rev. Lett.}\ }\textbf
  {\bibinfo {volume} {103}},\ \bibinfo {pages} {010602} (\bibinfo {year}
  {2009})}\BibitemShut {NoStop}%
\bibitem [{\citenamefont {Maes}(2017)}]{maes2017}%
  \BibitemOpen
  \bibfield  {author} {\bibinfo {author} {\bibfnamefont {C.}~\bibnamefont
  {Maes}},\ }\bibfield  {title} {\bibinfo {title} {Frenetic {{Bounds}} on the
  {{Entropy Production}}},\ }\href
  {https://doi.org/10.1103/PhysRevLett.119.160601} {\bibfield  {journal}
  {\bibinfo  {journal} {Phys. Rev. Lett.}\ }\textbf {\bibinfo {volume} {119}},\
  \bibinfo {pages} {160601} (\bibinfo {year} {2017})}\BibitemShut {NoStop}%
\bibitem [{\citenamefont {Maes}(2018)}]{maes2018}%
  \BibitemOpen
  \bibfield  {author} {\bibinfo {author} {\bibfnamefont {C.}~\bibnamefont
  {Maes}},\ }\href {https://doi.org/10.1007/978-3-319-67780-4} {\emph {\bibinfo
  {title} {Non-{{Dissipative Effects}} in {{Nonequilibrium Systems}}}}},\
  {{SpringerBriefs}} in {{Complexity}}\ (\bibinfo  {publisher} {{Springer
  International Publishing}},\ \bibinfo {address} {{Cham}},\ \bibinfo {year}
  {2018})\BibitemShut {NoStop}%
\bibitem [{\citenamefont {Maes}(2020)}]{maes2020}%
  \BibitemOpen
  \bibfield  {author} {\bibinfo {author} {\bibfnamefont {C.}~\bibnamefont
  {Maes}},\ }\bibfield  {title} {\bibinfo {title} {Frenesy: Time-symmetric
  dynamical activity in nonequilibria},\ }\href
  {https://doi.org/10.1016/j.physrep.2020.01.002} {\bibfield  {journal}
  {\bibinfo  {journal} {Physics Reports}\ }\textbf {\bibinfo {volume} {850}},\
  \bibinfo {pages} {1} (\bibinfo {year} {2020})}\BibitemShut {NoStop}%
\bibitem [{\citenamefont {Shiraishi}(2022)}]{shiraishi2022}%
  \BibitemOpen
  \bibfield  {author} {\bibinfo {author} {\bibfnamefont {N.}~\bibnamefont
  {Shiraishi}},\ }\bibfield  {title} {\bibinfo {title} {Time-symmetric current
  and its fluctuation response relation around nonequilibrium stalling
  stationary state},\ }\href {https://doi.org/10.1103/PhysRevLett.129.020602}
  {\bibfield  {journal} {\bibinfo  {journal} {Phys. Rev. Lett.}\ }\textbf
  {\bibinfo {volume} {129}},\ \bibinfo {pages} {020602} (\bibinfo {year}
  {2022})},\ \Eprint {https://arxiv.org/abs/2111.09477} {2111.09477}
  \BibitemShut {NoStop}%
\bibitem [{\citenamefont {Garrahan}(2017)}]{garrahan2017}%
  \BibitemOpen
  \bibfield  {author} {\bibinfo {author} {\bibfnamefont {J.~P.}\ \bibnamefont
  {Garrahan}},\ }\bibfield  {title} {\bibinfo {title} {Simple bounds on
  fluctuations and uncertainty relations for first-passage times of counting
  observables},\ }\href {https://doi.org/10.1103/PhysRevE.95.032134} {\bibfield
   {journal} {\bibinfo  {journal} {Phys. Rev. E}\ }\textbf {\bibinfo {volume}
  {95}},\ \bibinfo {pages} {032134} (\bibinfo {year} {2017})}\BibitemShut
  {NoStop}%
\bibitem [{\citenamefont {Di~Terlizzi}\ and\ \citenamefont
  {Baiesi}(2019)}]{diterlizzi2019}%
  \BibitemOpen
  \bibfield  {author} {\bibinfo {author} {\bibfnamefont {I.}~\bibnamefont
  {Di~Terlizzi}}\ and\ \bibinfo {author} {\bibfnamefont {M.}~\bibnamefont
  {Baiesi}},\ }\bibfield  {title} {\bibinfo {title} {Kinetic uncertainty
  relation},\ }\href {https://doi.org/10.1088/1751-8121/aaee34} {\bibfield
  {journal} {\bibinfo  {journal} {J. Phys. A: Math. Theor.}\ }\textbf {\bibinfo
  {volume} {52}},\ \bibinfo {pages} {02LT03} (\bibinfo {year} {2019})},\
  \Eprint {https://arxiv.org/abs/1809.06410} {1809.06410} \BibitemShut
  {NoStop}%
\bibitem [{\citenamefont {Vo}\ \emph {et~al.}(2022)\citenamefont {Vo},
  \citenamefont {Van~Vu},\ and\ \citenamefont {Hasegawa}}]{vo2022a}%
  \BibitemOpen
  \bibfield  {author} {\bibinfo {author} {\bibfnamefont {V.~T.}\ \bibnamefont
  {Vo}}, \bibinfo {author} {\bibfnamefont {T.}~\bibnamefont {Van~Vu}},\ and\
  \bibinfo {author} {\bibfnamefont {Y.}~\bibnamefont {Hasegawa}},\ }\bibfield
  {title} {\bibinfo {title} {Unified thermodynamic-kinetic uncertainty
  relation},\ }\href {https://doi.org/10.1088/1751-8121/ac9099} {\bibfield
  {journal} {\bibinfo  {journal} {J. Phys. A: Math. Theor.}\ }\textbf {\bibinfo
  {volume} {55}},\ \bibinfo {pages} {405004} (\bibinfo {year}
  {2022})}\BibitemShut {NoStop}%
\bibitem [{\citenamefont {Shiraishi}(2023{\natexlab{b}})}]{shiraishi2023a}%
  \BibitemOpen
  \bibfield  {author} {\bibinfo {author} {\bibfnamefont {N.}~\bibnamefont
  {Shiraishi}},\ }\href {https://doi.org/10.1007/978-981-19-8186-9} {\emph
  {\bibinfo {title} {An {{Introduction}} to {{Stochastic Thermodynamics}}:
  {{From Basic}} to {{Advanced}}}}},\ \bibinfo {series} {Fundamental
  {{Theories}} of {{Physics}}}, Vol.\ \bibinfo {volume} {212}\ (\bibinfo
  {publisher} {{Springer Nature Singapore}},\ \bibinfo {address}
  {{Singapore}},\ \bibinfo {year} {2023})\BibitemShut {NoStop}%
\bibitem [{\citenamefont {Harunari}\ \emph {et~al.}(2022)\citenamefont
  {Harunari}, \citenamefont {Dutta}, \citenamefont {Polettini},\ and\
  \citenamefont {Rold{\'a}n}}]{harunari2022}%
  \BibitemOpen
  \bibfield  {author} {\bibinfo {author} {\bibfnamefont {P.~E.}\ \bibnamefont
  {Harunari}}, \bibinfo {author} {\bibfnamefont {A.}~\bibnamefont {Dutta}},
  \bibinfo {author} {\bibfnamefont {M.}~\bibnamefont {Polettini}},\ and\
  \bibinfo {author} {\bibfnamefont {{\'E}.}~\bibnamefont {Rold{\'a}n}},\
  }\bibfield  {title} {\bibinfo {title} {What to {{Learn}} from a {{Few Visible
  Transitions}}' {{Statistics}}?},\ }\href
  {https://doi.org/10.1103/PhysRevX.12.041026} {\bibfield  {journal} {\bibinfo
  {journal} {Phys. Rev. X}\ }\textbf {\bibinfo {volume} {12}},\ \bibinfo
  {pages} {041026} (\bibinfo {year} {2022})}\BibitemShut {NoStop}%
\bibitem [{\citenamefont {Baiesi}\ \emph {et~al.}(2023)\citenamefont {Baiesi},
  \citenamefont {Falasco},\ and\ \citenamefont {Nishiyama}}]{baiesi2023}%
  \BibitemOpen
  \bibfield  {author} {\bibinfo {author} {\bibfnamefont {M.}~\bibnamefont
  {Baiesi}}, \bibinfo {author} {\bibfnamefont {G.}~\bibnamefont {Falasco}},\
  and\ \bibinfo {author} {\bibfnamefont {T.}~\bibnamefont {Nishiyama}},\
  }\href@noop {} {\bibinfo {title} {Effective estimation of entropy production
  with lacking data}} (\bibinfo {year} {2023})\BibitemShut {NoStop}%
\bibitem [{\citenamefont {Vu}\ and\ \citenamefont
  {Saito}(2023)}]{vanvu2023geometric}%
  \BibitemOpen
  \bibfield  {author} {\bibinfo {author} {\bibfnamefont {T.~V.}\ \bibnamefont
  {Vu}}\ and\ \bibinfo {author} {\bibfnamefont {K.}~\bibnamefont {Saito}},\
  }\href@noop {} {\bibinfo {title} {Geometric characterization for cyclic heat
  engines far from equilibrium}} (\bibinfo {year} {2023}),\ \Eprint
  {https://arxiv.org/abs/2305.06219} {arXiv:2305.06219} \BibitemShut {NoStop}%
\bibitem [{\citenamefont {Nishiyama}\ and\ \citenamefont
  {Hasegawa}(2023)}]{nishiyama2023}%
  \BibitemOpen
  \bibfield  {author} {\bibinfo {author} {\bibfnamefont {T.}~\bibnamefont
  {Nishiyama}}\ and\ \bibinfo {author} {\bibfnamefont {Y.}~\bibnamefont
  {Hasegawa}},\ }\bibfield  {title} {\bibinfo {title} {Upper bound for entropy
  production in {{Markov}} processes},\ }\href
  {https://doi.org/10.1103/PhysRevE.108.044139} {\bibfield  {journal} {\bibinfo
   {journal} {Phys. Rev. E}\ }\textbf {\bibinfo {volume} {108}},\ \bibinfo
  {pages} {044139} (\bibinfo {year} {2023})}\BibitemShut {NoStop}%
\bibitem [{\citenamefont {Hasegawa}(2023)}]{hasegawa2023}%
  \BibitemOpen
  \bibfield  {author} {\bibinfo {author} {\bibfnamefont {Y.}~\bibnamefont
  {Hasegawa}},\ }\href@noop {} {\bibinfo {title} {Thermodynamic correlation
  inequality}} (\bibinfo {year} {2023}),\ \Eprint
  {https://arxiv.org/abs/2301.03060} {arXiv:2301.03060 [quant-ph]} \BibitemShut
  {NoStop}%
\bibitem [{\citenamefont {van Kampen}(1992)}]{kampen1992}%
  \BibitemOpen
  \bibfield  {author} {\bibinfo {author} {\bibfnamefont {N.~G.}\ \bibnamefont
  {van Kampen}},\ }\href@noop {} {\emph {\bibinfo {title} {Stochastic Processes
  in Physics and Chemistry}}},\ \bibinfo {edition} {rev. and enl. ed}\ ed.\
  (\bibinfo  {publisher} {{North-Holland}},\ \bibinfo {address} {{Amsterdam}},\
  \bibinfo {year} {1992})\BibitemShut {NoStop}%
\bibitem [{\citenamefont {Schnakenberg}(1976)}]{schnakenberg1976}%
  \BibitemOpen
  \bibfield  {author} {\bibinfo {author} {\bibfnamefont {J.}~\bibnamefont
  {Schnakenberg}},\ }\bibfield  {title} {\bibinfo {title} {Network theory of
  microscopic and macroscopic behavior of master equation systems},\ }\href
  {https://doi.org/10.1103/RevModPhys.48.571} {\bibfield  {journal} {\bibinfo
  {journal} {Rev. Mod. Phys.}\ }\textbf {\bibinfo {volume} {48}},\ \bibinfo
  {pages} {571} (\bibinfo {year} {1976})}\BibitemShut {NoStop}%
\bibitem [{\citenamefont {Seifert}(2012)}]{seifert2012}%
  \BibitemOpen
  \bibfield  {author} {\bibinfo {author} {\bibfnamefont {U.}~\bibnamefont
  {Seifert}},\ }\bibfield  {title} {\bibinfo {title} {Stochastic
  thermodynamics, fluctuation theorems and molecular machines},\ }\href
  {https://doi.org/10.1088/0034-4885/75/12/126001} {\bibfield  {journal}
  {\bibinfo  {journal} {Rep. Prog. Phys.}\ }\textbf {\bibinfo {volume} {75}},\
  \bibinfo {pages} {126001} (\bibinfo {year} {2012})}\BibitemShut {NoStop}%
\bibitem [{\citenamefont {Peliti}\ and\ \citenamefont
  {Pigolotti}(2021)}]{peliti2021}%
  \BibitemOpen
  \bibfield  {author} {\bibinfo {author} {\bibfnamefont {L.}~\bibnamefont
  {Peliti}}\ and\ \bibinfo {author} {\bibfnamefont {S.}~\bibnamefont
  {Pigolotti}},\ }\href@noop {} {\emph {\bibinfo {title} {Stochastic
  Thermodynamics: An Introduction}}}\ (\bibinfo  {publisher} {{Princeton
  University Press}},\ \bibinfo {address} {{Princeton : Oxford}},\ \bibinfo
  {year} {2021})\BibitemShut {NoStop}%
\bibitem [{\citenamefont {Kettling}\ \emph {et~al.}(1998)\citenamefont
  {Kettling}, \citenamefont {Koltermann}, \citenamefont {Schwille},\ and\
  \citenamefont {Eigen}}]{kettling1998}%
  \BibitemOpen
  \bibfield  {author} {\bibinfo {author} {\bibfnamefont {U.}~\bibnamefont
  {Kettling}}, \bibinfo {author} {\bibfnamefont {A.}~\bibnamefont
  {Koltermann}}, \bibinfo {author} {\bibfnamefont {P.}~\bibnamefont
  {Schwille}},\ and\ \bibinfo {author} {\bibfnamefont {M.}~\bibnamefont
  {Eigen}},\ }\bibfield  {title} {\bibinfo {title} {Real-time enzyme kinetics
  monitored by dual-color fluorescence cross-correlation spectroscopy},\ }\href
  {https://doi.org/10.1073/pnas.95.4.1416} {\bibfield  {journal} {\bibinfo
  {journal} {Proc. Natl. Acad. Sci. U.S.A.}\ }\textbf {\bibinfo {volume}
  {95}},\ \bibinfo {pages} {1416} (\bibinfo {year} {1998})}\BibitemShut
  {NoStop}%
\bibitem [{\citenamefont {Bacia}\ \emph {et~al.}(2006)\citenamefont {Bacia},
  \citenamefont {Kim},\ and\ \citenamefont {Schwille}}]{bacia2006}%
  \BibitemOpen
  \bibfield  {author} {\bibinfo {author} {\bibfnamefont {K.}~\bibnamefont
  {Bacia}}, \bibinfo {author} {\bibfnamefont {S.~A.}\ \bibnamefont {Kim}},\
  and\ \bibinfo {author} {\bibfnamefont {P.}~\bibnamefont {Schwille}},\
  }\bibfield  {title} {\bibinfo {title} {Fluorescence cross-correlation
  spectroscopy in living cells},\ }\href {https://doi.org/10.1038/nmeth822}
  {\bibfield  {journal} {\bibinfo  {journal} {Nat Methods}\ }\textbf {\bibinfo
  {volume} {3}},\ \bibinfo {pages} {83} (\bibinfo {year} {2006})}\BibitemShut
  {NoStop}%
\bibitem [{\citenamefont {Shiraishi}(2021)}]{shiraishi2021a}%
  \BibitemOpen
  \bibfield  {author} {\bibinfo {author} {\bibfnamefont {N.}~\bibnamefont
  {Shiraishi}},\ }\bibfield  {title} {\bibinfo {title} {Optimal thermodynamic
  uncertainty relation in {{Markov}} jump processes},\ }\href
  {https://doi.org/10.1007/s10955-021-02829-8} {\bibfield  {journal} {\bibinfo
  {journal} {J. Stat. Phys.}\ }\textbf {\bibinfo {volume} {185}},\ \bibinfo
  {pages} {19} (\bibinfo {year} {2021})}\BibitemShut {NoStop}%
\bibitem [{\citenamefont {Uhl}\ and\ \citenamefont {Seifert}(2019)}]{uhl2019}%
  \BibitemOpen
  \bibfield  {author} {\bibinfo {author} {\bibfnamefont {M.}~\bibnamefont
  {Uhl}}\ and\ \bibinfo {author} {\bibfnamefont {U.}~\bibnamefont {Seifert}},\
  }\bibfield  {title} {\bibinfo {title} {Affinity-dependent bound on the
  spectrum of stochastic matrices},\ }\href
  {https://doi.org/10.1088/1751-8121/ab3a7a} {\bibfield  {journal} {\bibinfo
  {journal} {J. Phys. A: Math. Theor.}\ }\textbf {\bibinfo {volume} {52}},\
  \bibinfo {pages} {405002} (\bibinfo {year} {2019})}\BibitemShut {NoStop}%
\bibitem [{\citenamefont {Dechant}(2023)}]{dechant2023}%
  \BibitemOpen
  \bibfield  {author} {\bibinfo {author} {\bibfnamefont {A.}~\bibnamefont
  {Dechant}},\ }\href@noop {} {\bibinfo {title} {Thermodynamic constraints on
  the power spectral density in and out of equilibrium}} (\bibinfo {year}
  {2023}),\ \Eprint {https://arxiv.org/abs/2306.00417} {arXiv:2306.00417
  [cond-mat.stat-mech]} \BibitemShut {NoStop}%
\bibitem [{\citenamefont {Kolchinsky}\ \emph {et~al.}(2024)\citenamefont
  {Kolchinsky}, \citenamefont {Ohga},\ and\ \citenamefont
  {Ito}}]{kolchinsky2024}%
  \BibitemOpen
  \bibfield  {author} {\bibinfo {author} {\bibfnamefont {A.}~\bibnamefont
  {Kolchinsky}}, \bibinfo {author} {\bibfnamefont {N.}~\bibnamefont {Ohga}},\
  and\ \bibinfo {author} {\bibfnamefont {S.}~\bibnamefont {Ito}},\ }\bibfield
  {title} {\bibinfo {title} {Thermodynamic bound on spectral perturbations,
  with applications to oscillations and relaxation dynamics},\ }\href
  {https://doi.org/10.1103/PhysRevResearch.6.013082} {\bibfield  {journal}
  {\bibinfo  {journal} {Phys. Rev. Research}\ }\textbf {\bibinfo {volume}
  {6}},\ \bibinfo {pages} {013082} (\bibinfo {year} {2024})}\BibitemShut
  {NoStop}%
\bibitem [{\citenamefont {Pietzonka}\ \emph
  {et~al.}(2016{\natexlab{a}})\citenamefont {Pietzonka}, \citenamefont
  {Barato},\ and\ \citenamefont {Seifert}}]{pietzonka2016a}%
  \BibitemOpen
  \bibfield  {author} {\bibinfo {author} {\bibfnamefont {P.}~\bibnamefont
  {Pietzonka}}, \bibinfo {author} {\bibfnamefont {A.~C.}\ \bibnamefont
  {Barato}},\ and\ \bibinfo {author} {\bibfnamefont {U.}~\bibnamefont
  {Seifert}},\ }\bibfield  {title} {\bibinfo {title} {Universal bound on the
  efficiency of molecular motors},\ }\href
  {https://doi.org/10.1088/1742-5468/2016/12/124004} {\bibfield  {journal}
  {\bibinfo  {journal} {J. Stat. Mech.}\ }\textbf {\bibinfo {volume} {2016}},\
  \bibinfo {pages} {124004} (\bibinfo {year} {2016}{\natexlab{a}})}\BibitemShut
  {NoStop}%
\bibitem [{\citenamefont {Pietzonka}\ \emph
  {et~al.}(2016{\natexlab{b}})\citenamefont {Pietzonka}, \citenamefont
  {Barato},\ and\ \citenamefont {Seifert}}]{pietzonka2016}%
  \BibitemOpen
  \bibfield  {author} {\bibinfo {author} {\bibfnamefont {P.}~\bibnamefont
  {Pietzonka}}, \bibinfo {author} {\bibfnamefont {A.~C.}\ \bibnamefont
  {Barato}},\ and\ \bibinfo {author} {\bibfnamefont {U.}~\bibnamefont
  {Seifert}},\ }\bibfield  {title} {\bibinfo {title} {Affinity- and
  topology-dependent bound on current fluctuations},\ }\href
  {https://doi.org/10.1088/1751-8113/49/34/34LT01} {\bibfield  {journal}
  {\bibinfo  {journal} {J. Phys. A: Math. Theor.}\ }\textbf {\bibinfo {volume}
  {49}},\ \bibinfo {pages} {34LT01} (\bibinfo {year}
  {2016}{\natexlab{b}})}\BibitemShut {NoStop}%
\bibitem [{\citenamefont {Van~Vu}\ and\ \citenamefont
  {Hasegawa}(2019)}]{vanvu2019}%
  \BibitemOpen
  \bibfield  {author} {\bibinfo {author} {\bibfnamefont {T.}~\bibnamefont
  {Van~Vu}}\ and\ \bibinfo {author} {\bibfnamefont {Y.}~\bibnamefont
  {Hasegawa}},\ }\bibfield  {title} {\bibinfo {title} {Uncertainty relations
  for underdamped {{Langevin}} dynamics},\ }\href
  {https://doi.org/10.1103/PhysRevE.100.032130} {\bibfield  {journal} {\bibinfo
   {journal} {Phys. Rev. E}\ }\textbf {\bibinfo {volume} {100}},\ \bibinfo
  {pages} {032130} (\bibinfo {year} {2019})}\BibitemShut {NoStop}%
\bibitem [{\citenamefont {Fu}\ and\ \citenamefont {Gingrich}(2022)}]{fu2022}%
  \BibitemOpen
  \bibfield  {author} {\bibinfo {author} {\bibfnamefont {R.-S.}\ \bibnamefont
  {Fu}}\ and\ \bibinfo {author} {\bibfnamefont {T.~R.}\ \bibnamefont
  {Gingrich}},\ }\bibfield  {title} {\bibinfo {title} {Thermodynamic
  uncertainty relation for {{Langevin}} dynamics by scaling time},\ }\href
  {https://doi.org/10.1103/PhysRevE.106.024128} {\bibfield  {journal} {\bibinfo
   {journal} {Phys. Rev. E}\ }\textbf {\bibinfo {volume} {106}},\ \bibinfo
  {pages} {024128} (\bibinfo {year} {2022})}\BibitemShut {NoStop}%
\bibitem [{\citenamefont {Van Der~Meer}\ \emph {et~al.}(2022)\citenamefont {Van
  Der~Meer}, \citenamefont {Ertel},\ and\ \citenamefont
  {Seifert}}]{vandermeer2022}%
  \BibitemOpen
  \bibfield  {author} {\bibinfo {author} {\bibfnamefont {J.}~\bibnamefont {Van
  Der~Meer}}, \bibinfo {author} {\bibfnamefont {B.}~\bibnamefont {Ertel}},\
  and\ \bibinfo {author} {\bibfnamefont {U.}~\bibnamefont {Seifert}},\
  }\bibfield  {title} {\bibinfo {title} {Thermodynamic {{Inference}} in
  {{Partially Accessible Markov Networks}}: {{A Unifying Perspective}} from
  {{Transition-Based Waiting Time Distributions}}},\ }\href
  {https://doi.org/10.1103/PhysRevX.12.031025} {\bibfield  {journal} {\bibinfo
  {journal} {Phys. Rev. X}\ }\textbf {\bibinfo {volume} {12}},\ \bibinfo
  {pages} {031025} (\bibinfo {year} {2022})}\BibitemShut {NoStop}%
\bibitem [{\citenamefont {Ghosal}\ and\ \citenamefont
  {Bisker}(2022)}]{ghosal2022}%
  \BibitemOpen
  \bibfield  {author} {\bibinfo {author} {\bibfnamefont {A.}~\bibnamefont
  {Ghosal}}\ and\ \bibinfo {author} {\bibfnamefont {G.}~\bibnamefont
  {Bisker}},\ }\bibfield  {title} {\bibinfo {title} {Inferring entropy
  production rate from partially observed {{Langevin}} dynamics under
  coarse-graining},\ }\href {https://doi.org/10.1039/D2CP03064K} {\bibfield
  {journal} {\bibinfo  {journal} {Phys. Chem. Chem. Phys.}\ }\textbf {\bibinfo
  {volume} {24}},\ \bibinfo {pages} {24021} (\bibinfo {year}
  {2022})}\BibitemShut {NoStop}%
\end{thebibliography}

\end{document}